\documentclass[universe,article,accept,pdftex,moreauthors]{article}

\usepackage{color}
\usepackage{amsmath}
\usepackage{amssymb}
\usepackage{mathtools}
\usepackage[normalem]{ulem}
\usepackage[utf8]{inputenc}

\newcommand\h[1]{\ensuremath{#1}}

\newcommand\tu[1]{\h{^{#1}}}
\newcommand\dn[1]{\h{ _{#1}}}
\newcommand\f[2]{\h{\frac{#1}{#2}}}
\newcommand\brf[2]{\h{\lrr{\f{#1}{#2}}}}
\newcommand\bsf[2]{\h{\lrs{\f{#1}{#2}}}}

\newcommand\dsf[2]{\h{\ds \f{#1}{#2}}}
\newcommand\colb[1]{\textcolor{black}{#1}}

\newcommand\eto[1]{\h{e^{ #1}}}
\newcommand\gu[2]{\h{g\tu{#1 #2}}}
\newcommand\gd[2]{\h{g\dn{#1 #2}}}

\newcommand\sps[1]{\h{^{(#1)}}} 
\newcommand\sbs[1]{\h{\uu{(#1)}}}

\newcommand\deud[2]{\h{\de\upd{#1}{#2}}}
\newcommand\dedd[2]{\h{\de_{#1,#2}}}

\newcommand\dobydo[2]{\h{\dsf{\partial #1}{\partial #2}}}

\newcommand\mtrx[2]{\h{\lrs{\ba{#1} #2 \ea}}}
\newcommand\lrr[1]{\h{\left( #1 \right)}}
\newcommand\lrs[1]{\h{\left[ \, #1 \, \right]}}
\newcommand\lrc[1]{\h{\left\{ #1 \right\}}}
\newcommand\lrb[1]{\h{\left| #1 \right|}}

\newcommand\lt[2]{\lim_{#1 \ra #2}}
\newcommand\ct[1]{(\ref{eqn#1})}

\newcommand\hs[1]{\h{\hspace{#1in}}}
\newcommand\vs[1]{  \ \vspace{#1in} }

\newcommand\ti[1]{\h{\tilde{#1}}}

\newcommand\uu[1]{_{_{#1}}}

\newcommand\leqn[1]{\label{eqn#1}}

\newcommand\crop[4]{}

\newcommand\fs[3]{{\fontsize{#1}{#2}\selectfont #3 }}
\newcommand\upd[2]{{\tu{#1}}\,\dn{#2}}
\newcommand\dnu[2]{{\dn{#1}}\,\tu{#2}}

\newcommand\scn[2]{\h{#1 \times 10^{#2}}}
\newcommand\pot[1]{\h{10^{#1}}}

\newcommand\potm[1]{\h{10^{-#1}}}  
\newcommand\potmo[1]{\h{10^{-1#1}}}

\newcommand\cfl[3]{\Gamma\tu #1\  \dn{#2#3}}

\newcommand\rmn[4]{{R\tu{#1}}\dn{#2#3#4}}

\newcommand\mb[1]{\mbox{#1}}

\newcommand\dowcr[3]{\uu{#1 #2, #3}} 
\newcommand\doswsr[3]{\dn{#1; #2; #3}} 
 
\newcommand\dosw[2]{\dn{#1;#2}} 
\newcommand\metd[3]{g\dowcr #1 #2 #3}

\def\ssth{\sin\sq(\th)}

\newcommand\fdn[2]{F\dn{#1#2}}

\newcommand\fdu[2]{F\dn{#1}\, \tu{#2}}

\def\bdf{B\dn f}
\def\ldf{l\dn f}

\newcommand\dowsr[3]{\dn{#1 #2 ; #3}}
 \def\z{\zeta}

 \def\rs{r\dn s}
 \newcommand\dso[1]{{}\dn{;#1}}

\title{Primordial Magnetogenesis from Killing Vector Fields}

\author{Nagabhushana Prabhu}
\date{Purdue University, West Lafayette, IN 47907}
\def\ds{\displaystyle}

\def\beq{\begin{equation}}
\def\eeq{\end{equation}}
\def\bi{\begin{itemize}}
\def\be{\begin{enumerate}}
\def\bc{\begin{center}}
\def\bt{\begin{tabular}}
\def\bd{\begin{description}}
\def\bq{\begin{quotation}}
\def\ba{\begin{array}}
\def\beas{\begin{eqnarray*}}
\def\bea{\begin{eqnarray}}
\def\bv{\begin{verbatim}}
\def\ev{\end{verbatim}}
\def\eea{\end{eqnarray}}
\def\eeas{\end{eqnarray*}}
\def\ea{\end{array}}
\def\eq{\end{quotation}}
\def\ed{\end{description}}
\def\ec{\end{center}}
\def\et{\end{tabular}}
\def\ee{\end{enumerate}}
\def\ei{\end{itemize}}

\def\d{\h{\cdot}}
\def\j{ & }
\def\ada{ \j \dfn \j}
\def\aea{ \j = \j}

\def\aba{ \j   \j}

\def\alea{ \j \leq \j }
\def\aga{ \j > \j }

\def\nn{\nonumber \\}

\def\a{\h{\alpha}}
\def\b{\h{\beta}}
\def\g{\h{\gamma}}
\def\th{\h{\theta}}
\def\de{\h{\delta}}

\def\k{\h{\kappa}}
\def\om{\h{\omega}}
\def\m{\h{\mu}}
\def\n{\h{\nu}}
\def\sg{\h{\sigma}}

\def\rh{\h{\rho}}
\def\e{\h{\epsilon}}
\def\l{\h{\lambda}}
\def\L{\h{\Lambda}}
\def\vp{\h{\varphi}}

\def\ez{\h{\e_0}}

\def\s{\section}
\def\ss{\subsection}

\def\sz{\h{\uu 0}}
\def\so{\h{\uu 1}}

\def\uz{\h{^ 0}}

\def\sq{\h{^ 2}}
\def\cu{\h{^ 3}}
\def\uv{\h{^ 5}}

\def\upn{\h{^ n}}

\def\upm{\h{^\mu}}
\def\dnm{\h{_\mu}}
\def\upa{\h{^\a}}
\def\dna{\h{_\a}}
\def\upb{\h{^\b}}
\def\dnb{\h{_\b}}
\def\upn{\h{^\nu}}
\def\dnn{\h{_\nu}}

\def\dns{\h{_\sg}}

\def\dnl{\h{_\l}}

\def\dnr{\h{_\rh}}

\def\tuh{\h{\tu{\f12}}}

\def\inv{\h{^{-1}}}
\def\dmn{\h{_{\m\n}}}
\def\umn{\h{^{\m\n}}}
\def\dab{\h{_{\a\b}}}
\def\uab{\h{^{\a\b}}}
\def\drs{\h{_{\rho\sigma}}}
\def\urs{\h{^{\rho\sigma}}}
\def\sk{\h{\dn k}}

\def\tf{\h{^4}}

\def\pmd{\ensuremath{\pl_\mu}}
\def\pmu{\ensuremath{\pl^\mu}}

\def\pnd{\ensuremath{\pl_\nu}}

\def\ra{\h{\rightarrow}\, }

\def\pl{\partial}

\def\dfx{d\tu 4 x}

\def\uf{\tu 4}
\def\uv{\tu 5}

 \def\sep{\vspace{12pt} \hrule width \hsize \kern 1mm \hrule width \hsize height 2pt \vspace{12pt} }
 
 \def\cls{\h{{\cal S}}}

 \def\clr{\h{{\cal R}}}
 
\def\cla{\h{{\cal A}}}

\def\clf{\h{{\cal F}}}

\def\clj{\h{{\cal J}}}
\def\cll{\h{{\cal L}}}

\def\clg{\h{{\cal G}}}

\def\deq{\h{\ = \ }}

\def\beass{\begin{small}\beas}
\def\eeass{\eeas\end{small}}
 \def\beasm{\begin{small}\bea}
\def\eeasm{\eea\end{small}}
 \def\beafns{\begin{footnotesize}\bea}
\def\eeafns{\eea\end{footnotesize}}

\def\rtd{\right.}
\def\ltd{\left.}

 \def\clr{\h{{\cal R}}}
 
\def\cla{\h{{\cal A}}}

\def\clf{\h{{\cal F}}}

\def\cll{\h{{\cal L}}}

\def\clg{\h{{\cal G}}}

\def\dfn{\h{:=}}
\def\me{m\uu e}

\newcommand\mk[2]{\tu{#1}{}\dn{#2}} 
 \newcommand\km[2]{\dn{#1}{}\tu{#2}} 
 \newcommand\kk[2]{\dn{#1#2}} 
 \newcommand\mm[2]{\tu{#1#2}} 
 \newcommand\nk[1]{\nabla\dn{#1}}
\begin{document}
\maketitle
\begin{abstract}
Papapetrou showed that the covariant derivative of a Killing vector field  satisfies Maxwell's equations in vacuum \cite{papa}.  We extend Papapetrou's result to show that the covariant derivative of a Killing vector field satisfies Maxwell's equations  in   non-vacuum backgrounds as well, if we allow  electromagnetic currents of purely geometric origin.  
We postulate that  every Killing vector field gives rise to a physical electromagnetic field and, in a non-vacuum background, a physical electromagnetic current---hereafter called {\it Killing electromagnetic field} and {\it Killing electromagnetic 
current}, respectively.
We show that the Killing 
electromagnetic field of the flat 
FLRW (Friedmann-Lema$\hat{\mbox{i}}$tre-Robertson-Walker) universe comprises a Killing magnetic field and a rotational Killing electric field; 
we derive an upper bound
on the Killing magnetic field
and find that the upper bound is consistent with the current 
upper bound on the cosmic magnetic field.  
We 
show that the time-like Killing vector of the Schwarzschild spacetime gives rise to a 
radial Killing electric field. We also show that in the weak field 
regime---and far from the
matter distribution---the back reaction of the radial Killing electric field changes the
Schwarzschild metric to Reissner-Nordstr$\ddot{\mbox{o}}$m metric; our calculation yields the
converse of Wald's result \cite{wald}.   Next, drawing upon Rainich's  
work on Rainich-Riemann manifolds \cite{rain}, we  discuss the etiological 
question of how a physical electromagnetic field can arise out of 
geometry; we also argue that detection of the Killing electric field of 
flat FLRW spacetime may be within current experimental reach.
Finally, we  discuss  the relevance of  
Killing electromagnetic currents and the aforementioned    
transmutation of Schwarzschild spacetime to Reissner-Nordstrom
spacetime, to 
Misner and Wheeler's 
program of realizing ``charge without charge'' \cite{miwh}.
\end{abstract}
\section{Introduction}
\noindent The origin of the cosmic magnetic field---a weak magnetic field that   
 pervades  the intergalactic region---remains a mystery 
 \cite{dola,kron,eaku,bfsi,grru}. The  current
observational bounds on the strength of the cosmic magnetic field (CMF)---also 
called the intergalactic or extragalactic magnetic field---and 
the sourcing mechanisms that have been proposed to explain its origin are summarized 
in 
\colb{\S}\ref{sec:cmf}.  Given the vastness of the intergalactic region and the relatively sparse 
distribution of \colb{the} galaxies that punctuate it \cite{gam2}, it is believed that the 
ubiquitous CMF likely did not originate from any  post-recombination 
astrophysical process(es) \cite{bfsi,grru}, but instead\colb{,}
 is likely of primordial origin and was generated by a mechanism that was operative over the entire
 vastness of intergalactic region \cite{dola,bfsi,grru,subr}.  
 
A second, possibly related, enigma can be stated, using Gamow's words \cite{gamo}, 
as the following question: 
why is it that ``\ldots {\it all successive degrees of accumulation of matter, such 
as planets, stars and galaxies, are found in the state of more or less rapid 
axial rotation\ldots}''?  In 
\colb{\S}\ref{room} we mention a related  
conjecture \colb{of Gamow's} that does not resolve but rather amplifies the enigma.
We also summarize 
the follow-up work on Gamow's conjecture, the 
\colb{current constraints on } vorticity in the universe\colb{,  and also previous attempts at explaining} the origin of 
axial rotations of galaxies.  

\colb{Against the backdrop of the above two enigmas, we look at}
an intriguing observation that Papapetrou made in
1966 \cite{papa}.  Specifically,  Papapetrou 
\colb{showed} that the covariant derivative 
of a Killing vector field\colb{,} of a spacetime\colb , resembles the electromagnetic field strength tensor\colb , and  satisfies the  
Maxwell's equations in vacuum 
\cite{papa}.  Wald used Papapetrou's observation  to derive the electromagnetic field around a Kerr black hole 
placed in a uniform magnetic field that is parallel to the black hole's
axis of rotation \cite{wald}.   Wald's construction has been used  
in the 
study of Kerr black holes by a number of authors \cite{ruff,frwi,maie,skks,spaj}.  
Most 
notably, Blandford and Znajek used Wald's construction to propose the so-called
BZ mechanism, which suggests that the astrophysical jets of a Kerr black hole are powered 
by its spin energy \cite{blzj}. 

\colb{We extend Papapetrou's result to show that the covariant derivative of a Killing vector field satisfies Maxwell's equations 
even in  non-vacuum backgrounds, if we allow  electromagnetic
currents of purely geometric origin; see Appendix \ref{app1}.} 


\colb{Guided by the results mentioned in the two previous paragraphs, we propose the following postulates.} 

\vs{-0.25}

\colb{
\bea
\mbox{Postulate 1: \parbox[t]{4.45in}{\it  Every Killing vector field gives rise to a   {\it physical} electromagnetic field whose field strength tensor  is proportional to the covariant derivative of the
Killing vector field.}} 
 \qquad\leqn{p1}\\
 \mbox{Postulate 2: \parbox[t]{4.45in}{\it  In non-vacuum background, every Killing vector field  gives rise to a physical electromagnetic current.}} 
 \qquad\leqn{p2}
\eea}
%
%
\colb{The electromagnetic field and the electromagnetic current---mentioned in Postulates 1 and 2---will be called the {\it Killing electromagnetic field} (KEF) and {\it Killing electromagnetic current} (KEC) respectively. We
will refer to Postulates 1 and 2 as KEF and KEC postulates, hereafter. }
%
%


\colb{
At the outset, a natural criticism of the postulates 
is that the postulated emergence of physical  electromagnetic 
field and current out of geometry (Killing vectors) 
seems far-fetched.
{\it What is the mechanism---the manner of causation---by which
Killing symmetries, which are geometric structures, give rise to
physical, measurable fields and currents?}
}

\colb{
Less than a decade after the formulation of general relativity,
Rainich \cite{rain} published a seminal calculation that, effectively, dissolved
the distinction between source-free electromagnetism and the 
geometry of spacetime. Summarized in \S \ref{mtfrt}, 
Rainich's  calculation---which  also served as the foundation 
for Misner and Wheeler's
work \cite{miwh}---shows how source-free electromagnetism
`emerges' from geometry.   
Rainich's calculation  suggests a conjectural explanation of how 
KEF and KEC can `emerge' from Killing vectors, and why the 
KEF and KEC postulates may not be very far-fetched.
 Admittedly, the postulates
are speculative conjectures until they are experimentally verified or falsified, and in \S\ref{falsifiability},  we  
argue that falsifying
the KEF postulate may be within current experimental reach.}


\colb{Predicated on the   validity of the KEF postulate, 
in \S\ref{fflrw} and \S\ref{schwarz}  we} 
explore \colb{its} consequences 
in two spacetimes---the flat FLRW   
spacetime, which is believed to describe the large scale structure 
of our universe \cite{plan}, and the Schwarzschild spacetime.

We show that three of the Killing vectors of the flat FLRW spacetime generate a \colb{Killing} 
 magnetic field that pervades the spacetime. Self-consistency 
constraint 
enables us to derive an upper bound on the strength of the \colb{Killing} magnetic field. We  
find that the derived upper bound 
is consistent with the known  bounds on the strength of the CMF.  
Three other Killing vectors of the flat FLRW spacetime  generate
\colb{a Killing} electric field, which has a nonzero \colb{`curl'}\colb{;
the} rotational 
Killing electric field 
raises the \colb{speculative} question of 
whether  {\it the \colb{Killing} electric field
 may have seeded  rotation of primordial\colb{,}  charged matter} and whether {\it it  
continues to power   
the ``axial rotation'' of charged matter today.}  
\colb{
The answer to the preceding question hinges
on 
the strength of the Killing electromagnetic fields in
flat FLRW spacetime; we discuss  bounds on the strengths of 
Killing electromagnetic fields in flat FLRW and Schwarzschild 
spacetimes in~\S \ref{fflrw} and \S \ref{schwarz}.}

\colb{Finally}, we explore the consequences of  \colb{KEF} postulate   in the Schwarzschild
spacetime.  The timelike Killing vector of the Schwarzschild spacetime generates
a radial \colb{Killing} electric field.   We derive the back reaction of the radial Killing electric
field on the Schwarzschild spacetime in the weak field regime
 and find that\colb{,} far from the distribution of matter\colb{,}  the back reaction   
changes  the Schwarzschild metric to   Reissner-Nordstr$\ddot{\mbox{o}}$m 
metric.  Thus, \colb{if the KEF postulate   is valid  then} an observer far from the spherical distribution of matter
concludes 
that the spacetime is described by the Reissner-Nordstr$\ddot{\mbox{o}}$m metric\colb{;  in \S \ref{cwc},  
we discuss the implications of the above transmutation of the 
metric by   Killing electric field.}

The paper is organized as follows. In 
\colb{\S  \ref{sec:cmf}} we summarize the known bounds on
the CMF and the mechanisms that have been proposed to explain its origin.  \colb{In \S\ref{room}}, we
summarize  highlights of  previous work on the puzzle of ubiquitous 
vorticity in accumulations of matter.  In Appendix \ref{app1} we extend Papapetrou's derivation to \colb{non-vacuum background and to 
conformal Killing vectors.}
In 
\colb{\S  \ref{fflrw}}, we 
derive the Killing magnetic field and the Killing electric field generated by the Killing vectors of
the flat FLRW spacetime, and study their \colb{conjectured} cosmological implications.  In 
\colb{\S \ref{schwarz}}, we
derive the radial Killing electric field emerging from the time-like Killing vector of Schwarzschild
spacetime, and show that\colb{,} in the weak field 
regime\colb{,}  and far from the distribution of matter\colb{,}  the Killing electric field changes the Schwarzschild 
metric to a Reissner-Nordstr$\ddot{\mbox{o}}$m metric.  \colb{In 
 \S  \ref{gauge} we discuss,   the gauge invariance of   KEF.  In 
 \S \ref{plausibility} we discuss the plausibility of the KEF and KEC postulates.
 In \S \ref{falsifiability} we argue that falsifying the KEF postulate may be 
 within current experimental reach.  
In \S \ref{mtfrt} we draw upon Rainich's results to provide a conjectural
description of how KEF and KEC could  `emerge' from Killing vectors.
 In \S \ref{cwc}, we discuss the transmutation of the Schwarzschild metric,
by its Killing electric field, into  Reissner-Nordstr$\ddot{\mbox{o}}$m
metric as well as the postulated KEC, in the context of Misner and Wheeler's results on ``charge without charge''. In \S \ref{mfkv}, we summarize the main features of
KEF and KEC  that distinguish magnetogenesis sourced by Killing vectors from
previous proposals for primordial magnetogenesis.}
In 
\colb{\S}\ref{future} we discuss some directions for future work.
\section{Background}\label{back}
 \noindent  In the first subsection below, we summarize the \colb{main}   
 bounds on the CMF and the
mechanisms that have been proposed to explain its origin. 
In the second subsection, we summarize the previous work on understanding 
the ``axial rotation'' observed in all accumulations of matter up to galactic 
length scales.
\subsection{Cosmic Magnetic Field}\label{sec:cmf}
 \noindent  A homogeneous CMF affects the  
isotropy of the cosmic microwave background (CMB) radiation.  
Hence the known bounds on CMB anisotropy can be used to 
derive an upper bound on the strength of a homogeneous CMF.
An upper bound of 
$\scn{3.4}{-9} (\Omega\sz h\dn{50}^2)\tuh$~G on  
the current strength of any primordial homogeneous magnetic was  derived 
by Barrow et al, based on an analysis of the CMB anisotropy data \cite{bfsi}; 
$\Omega\sz$ is the cosmological 
density parameter and $h\dn{50}$ is the current value of 
the Hubble constant in units of 50 km\d s\inv\d Mpc\inv.  
A homogeneous magnetic field Faraday-rotates polarized light
from distant sources, and hence the Faraday rotation can be 
used to estimate the strength of CMF.  Using rotation measure
data from 309 galaxies and quasars Vallee \cite{val1} derived an upper
bound of $\pot{-9} (\Omega\dn{IG} h\dn{100}/0.01)\inv$ G
on the strength of the CMF; $\Omega\dn{IG}$ is the ionized
gas density, as a fraction of the critical density, and $h\dn{100}$
is Hubble constant in units of 100  km\d s\inv\d Mpc\inv.  The 
recently discovered evidence of  ultrahigh-energy cosmic 
ray events emanating from the Perseus-Pisces supercluster was
used by Neronov et al. to derive an upper bound of about $\potm{10}$ 
G on the CMF \cite{nska}; this upper bound is arguably the strongest to date.
Finally, known constraints on big bang nucleosynthesis (BBN) also
impose an upper bound on the strength of the CMF, although the
upper bound derived from BBN is weaker than that derived 
from CMB anisotropy.  Therefore we do not discuss the upper
bound derived from BBN \cite{grru}, turning instead to 
the more interesting \colb{discussion of } 
lower bound on the strength of CMF. 

The electromagnetic cascade that is triggered when the TeV photons from 
blazars interact with the extragalactic photon background gives rise to secondary GeV photons.  The charged sector of the cascade couples to the CMF.  The nondetection of the secondary GeV photons by the Fermi/LAT telescope was used by Neronov and Vovk
\cite{nevo}  and Ackerman et al. \cite{acke} to establish a lower bound of $B\dn{CMF} \geq \scn
3 {-16}$ G on the CMF.  The order of magnitude of   Neronov and Vovk's
lower bound was reiterated subsequently by  Tavecchio et al. \cite{tave}  and Dolag et al \cite{dola}.  Dermer et al. \cite{derm}
argue that the lower bound would be lower and $B\dn{CMF} \ 
\gtrsim  \ \potmo 8$ G, if the TeV activity of the blazars is assumed
to last for less time than in the above reports.  Regardless, the 
GeV photon flux data from blazars provides a nonzero lower bound
on the strength of the CMF. 

We turn next to the sourcing mechanisms that may have given rise to 
the CMF.  
Thorne   
\cite{thor} conjectures
that the CMF must have come into existence at big bang.  Specifically, 
Thorne \cite{thor} writes ``{\it All attempts to understand how galactic
magnetic fields could have arisen since the big-bang creation of
the universe (assuming the big-bang theory to be correct) face 
difficulties, which seem to be insurmountable (see, e.g., Hoyle 1958).
The only way in which the existence of magnetic fields can be 
understood today is by assuming that they, like the matter in the 
universe, were created in the big bang.  Although, this suggestion
seems distasteful and {\it ad hoc} to many astrophysicists, the 
absence of any other explanation suggests that it be considered 
seriously (Zel'dovich 1964a).}''  We note that the Killing magnetic 
field described in this  paper is generated at the birth of the flat 
FLRW spacetime, and supports Thorne's conjecture 
\colb{in } that
the emergence of a magnetic field at big bang does not have to 
postulated\colb{;  it}  can be derived as an inescapable consequence of 
spacetime's Killing symmetries\footnote{Assuming that the flat 
FLRW spacetime emerged at big bang.}.

\colb{The vorticity in primordial plasma has  
been investigated as  possible source of magnetic field.  
Harrison \cite{harr} has suggested that if one 
assumes the existence of vorticity in the pre-recombination era then the 
differences   in the Thomson cross sections of the electrons and protons  
can give rise to a seed magnetic field.  If vorticity was not present
soon after big bang, however, and the assumptions of Helmholtz-Kelvin
circulation theorem hold, significant vorticity cannot emerge during 
subsequent 
evolution \cite{grru}. Rebhan \cite{rebh} has studied conditions under
which the Helmholtz-Kelvin theorem can be circumvented.  The constraints
on  vorticity in the early universe come from CMB anisotropy limits, using which 
Rebhan has also derived upper bound on the strength of the current
magnetic field.  It has also been argued  that the nonlinear 
evolution of primordial density perturbations can generate vorticity of 
ions and electrons at second order of perturbation \cite{mata,gose,taka,ichi}.
Siegel and Fry \cite{sifr} have argued that cosmological perturbations
in the pre-recombination era produce charge separations and currents
that produce seed magnetic fields.}

\colb{Vilenkin and Leahy\cite{vile} have argued that  parity violation in the 
electroweak theory gives rise to parity-violating currents, which 
can seed primordial magnetic field of sufficient strength. Brizard et al., \cite{bmwu}
have argued that in the presence of vorticity, plasma and neutrino-antineutrino
asymmetry, the neutrino-plasma interactions can generate magnetic fields.}

\colb{Inflation provides a kinematic means for exponentially 
increasing the coherence lengths of the magnetic fields that are produced
either before or during inflation.   However, if the universe is 
conformally flat---as FLRW spacetimes are---then, due to the conformal 
invariance of electromagnetism,  the magnetic field 
decays  too fast (as $1/a\sq$) with expansion to allow meaningfully large 
magnetic fields to exist at later times.  Therefore, previous researchers
have explored mechanisms to break the conformal invariance of 
electromagnetism.} 

\colb{Turner and Widrow\cite{tuwi} considered different models
in which   conformal invariance is broken, specifically, by coupling the electromagnetic 
field to gravity, or a scalar field that is not conformally invariant, or to
axions. 
Since the scale factor grows 
exponentially during inflation,
the strengths of the generated magnetic fields are sensitive to the
model parameters \cite{subr}.  For example,
Ratra \cite{ratr} proposed a model in which the electromagnetic field
is coupled to the inflaton (scalar) field; the parameters in the model
enable production of magnetic fields from $\lesssim \potm{65}$ G 
to $\gtrsim \potm{10}$ G.  Dolgov \cite{dolg} has considered the 
breaking of conformal invariance due to the trace anomaly in curved 
spacetime.}

\colb{As discussed in \cite{subr,mayo,fumu,kswa,dmru},
inflationary magnetogenesis has to address two problems: the back reaction 
problem and the strong coupling problem.
If the energy density of the produced electromagnetic 
field is too high---as would be the case for some values of parameters---then
the electromagnetic field has a nontrivial back reaction on magnetogenesis.  
Demozzi et al. \cite{dmru}
 argue that if the back reaction is assumed to be not strong enough to 
 subvert inflation, then the produced magnetic field would be too weak.  Also
 see \cite{dosi,dolg,ratr,lele,ggve,bptv,basa}. }
 
\colb{The second problem in inflationary magnetogenesis---called the strong coupling problem---is that, in order to get sufficient magnetic energy density at the present
time, the renormalized charge in the early stages of inflation may need
to be too large, putting the theory in the strong coupling regime where perturbative
analysis of electromagnetism does not hold.  Models that seek to circumvent the 
strong coupling problem have been proposed by several authors \cite{tasi,fjsl,camp},
including a proposal to use higher-dimensional scale factor \cite{assu}.
}

\colb{Campanelli \cite{camp} has shown that expected value of the 2-point
function of the magnetic field in de Sitter spacetime is scale invariant, without
explicitly breaking conformal invariance.  Also, see \cite{dmri,cam2} for 
discussion of the physical relevance of Campanelli's result. Gasperini  et al \cite{ggve}
have considered primordial magnetogenesis in string cosmology due to the 
coupling of electromagnetic field to a dilaton background.}

\colb{In addition to models that generate non-helical magnetic fields, models
that produce helical magnetic fields have also been proposed. For example, coupling
the $F\dmn \ti F\umn$ term to a pseudo-scalar field, such as an axion, produces
a helical magnetic field \cite{dhja}.  Despite the possible 
amplification of helical magnetic
fields by inverse cascade\cite{beol}, Durrer et al., \cite{dhja} conclude that the 
helical magnetic fields cannot produce the observed field strength 
\colb{over} 
the observed correlation lengths \cite{jdho}.}

\colb{Besides inflationary magnetogenesis and generation of magnetic fields
by primordial vorticity, the phase transitions in the early universe---particularly
the electroweak phase transition (EWPT)  and the QCD phase transition (QCDPT)---have also been
investigated as possible settings for magnetogenesis.  Although
experimental constraints have not definitively ruled out the 
possibility, it is believed
that   EWPT and QCDPT were not first order transitions 
\cite{dune}.  If the transitions were of first 
order, then bubbles of new phase formed and grew in a sea of old phase.  Baym et al. \cite{bbmc}
argue that magnetic fields of sufficient strength can be generated in a  
first order EWPT by shock waves caused by the expansion of bubbles of 
the new phase.  Brandenburg and Subramanian \cite{brsu} argue that the
turbulence caused by collision of bubbles could amplify magnetic fields
due to dynamo action.  Vachaspati \cite{vach} proposed that the gradients in the Higgs
field---arising from bubbles of the new phase having different vacuum expectation
values---could give rise to magnetic fields with strengths of the order of $\pot{23}$ 
G.  The mechanism proposed by Vachaspati have been further analyzed by 
Davidson \cite{davi} and Grasso and Riotto \cite{grri}.  The conclusions
of Vachaspati \cite{vach} and Grasso and Riotto \cite{grri} were numerically
confirmed by Diaz-Gil et al., \cite{diaz1,diaz2}.  Numerical simulations of 
the mechanism suggested by Vachaspati have also been reported in 
\cite{cfva,stjo,sjkh}. For suggested links between magnetogenesis and 
baryogenesis during EWPT, and numerical tests of the suggestion
 see \cite{cfva} and \cite{v2001}. }
 
 \colb{Similarly, if the QCDPT was of first order, the expanding bubbles of hadronic
 phase release latent heat in the form of shock waves into the quark-gluon
 plasma\cite{qlsp}.  The difference in the responses of the positive and 
 negative charges to the shock waves, Quashnock et al. argued, could 
 produce magnetic field of the order of 5 G.  Cheng and Olinto \cite{chol} argue that the currents produced at the interface between the hadronic phase and
 the quark-gluon plasma can produce magnetic fields of up to $\pot 8$ G.}
 
 \colb{Boyarsky et al. \cite{brsh}
 have suggested that the parity violation in the weak force  and the chiral 
 anomaly, generate a long-range magnetic field in the ground state of the 
 Standard Model.  Boyarsky et al. \cite{bfru} have argued that 
 the coherence scale of the above magnetic field could increase due to the
 so-called $\a$-effect.  For other work on helicity generation stemming
 from parity violation in the weak force see \cite{josh,fica,seso,ssva}. }
 
 \colb{Finally, it has been proposed that the CMF could also be seeded in the post-recombination era by 
outflows from magnetized galaxies into the intergalactic void \cite{kron,rees,dalo,enss,fulo}.  
For reviews of previous work on primordial magnetogenesis, see 
\cite{subr,kkts,grru,dune,vac2,widr,vall}.
 }
 
 \vspace{-12pt}

\subsection{Rotation of Observable Matter}\label{room}
\noindent In 1946 Gamow called the ``axial rotation'' of ``all successive
degrees of accumulation of matter'', ``One of the most mysterious
results of the astronomical studies of the universe'' \cite{gamo}.   
He then sought to expand the scope of the above ``mystery''  by asking, if 
in addition to the observed rotations up to the scale of galaxies,
``whether it is not 
possible to assume that {\it all matter in the visible universe
is in a state of general rotation around some centre located far 
beyond the reach of our telescopes?}'' 
G\"odel, who discovered the rotating universe solution to 
Einstein's equations a few years later \cite{gode}, refers to Gamow's
observation in his lecture in 1949 \cite{god2}. 

Much work has been done both to follow up on Gamow's hypothesis 
{\it that all matter is rotating about some cosmic centre}, as well as 
on vorticity in the universe in general.  For example,  Li observed that
Gamow's hypothesis, if valid, can explain galactic rotations as the 
effect of Coriolis force on radially collapsing matter in the course of galaxy 
formation \cite{lxli}.   Li  showed that Gamow's hypothesis can also
explain the empirically
observed relation \cite{bros,ozer,bubu,trim} between the angular 
momentum  $J$ and  mass $M$ of galaxies 
\beas
J\propto M\tu{5/3}.
\eeas 

Whereas Gamow hypothesized that all visible matter may be in a state of rotation,
Hawking \cite{hawk} and Hawking and Collins \cite{haw2} considered smaller 
scales of about 100 Mpc and derived limits on vorticity from the CMB data in various spatially 
homogeneous models of the universe.   Other upper bounds
on vorticity have been derived in \cite{elli,ell2,ell3,barr,jaff,jaf2,such,saad}. 
 
Gamow's hypothesis and the follow-up work on it focus on the length scales up to which
visible matter may be rotating. On the other hand, the aforementioned
 bounds on vorticity focus on the rate of
rotation of visible matter.  Neither line of inquiry, however, addresses the 
``mystery'' of {\it why} all visible matter   
is in a state of axial rotation.
In other words, a fundamental question is: {\it what, possibly primordial,
mechanism is responsible for injecting nonzero angular momentum into 
all accumulations of matter---at least up to the length scales of     
galaxies?}

As mentioned above, Li \cite{lxli} argues that Gamow's global rotation would 
explain the origin of rotation of galaxies. One concern with Li's explanation is 
that it would imply anisotropy in the axes of rotation of 
galaxies the existence and extent of which is still being debated \cite{parn,flin,godl}.  Hoyle \cite{hoy2} suggested that the tidal torques from
neighboring proto-accumulations of matter could be responsible for seeding
rotations in proto-galaxies.  Hoyle's suggestion was investigated further by 
Peebles \cite{peeb}. von Weizsacker \cite{vonw} and Gamow \cite{gam2} 
on the other hand argued that the rotations of galaxies could be seeded
by primordial turbulence, although Gamow notes that ``\ldots it is,  however,
difficult to see how such a (turbulent) motion could originate in a uniformly
expanding homogeneous material.''  Finally, Chernin \cite{cher} and Doroshkevich
\cite{doro} have suggested that vortices in primordial matter---arising from
shock waves generated by gravitational instability---as a possible explanation 
for the origin of rotations in galaxies. 
\s{{{Killing Electromagnetic Field of Flat FLRW Spacetime}}}\label{fflrw}
\noindent Observational evidence strongly suggests that   
the universe is described by the flat FLRW metric \cite{plan}.   
The metric of flat FLRW spacetime can be written as   
\bea
ds\sq \aea c\sq dt\sq - a(t)\sq (dx\sq + dy\sq + dz\sq) \leqn{fflrw}
\eea
where $a(t)$ is the scale parameter, and $x,y,z$\colb{,} the comoving spatial coordinates. 
The homogeneity of spatial hypersurface allows us to choose the 
spatial origin to be at an arbitrary point on the hypersurface.  

It is easily verified that
\fs{9}{11}{
\bea 
\lrr{K\sps 1}\upm \aea \brf 1 \ldf (0,0,-z,y), \nn      
\lrr{K\sps 2}\upm \aea \brf 1 \ldf (0,z,0, -x), \nn 
\lrr{K\sps 3}\upm \aea \brf 1 \ldf (0,-y,x,0),  
\leqn{kvs} 
\eea 
}
are contravariant Killing vectors of the flat FLRW metric \ct{fflrw}.  The   
constant\footnote{\label{fnt2} $\ldf \deq \dsf h {\me c} \sim \scn{2.4}{-12} m$. } 
$\ldf$  
 has the dimension of length and makes the covariant Killing 
vectors $(K\sps i)\colb{\upm}$ dimensionless. 
 
We define the  {\it electromagnetic field strength tensor} (EFST)  corresponding
to the  vector $K\sps i, i = 1,2,3$ as 
\bea
F\dmn\lrs{K\sps i}  \ada \zeta \colb{\lrr{K\sps i \dosw \n\m - K\sps i\dosw \m\n}}, 
\qquad \zeta \dfn \bdf \ldf \leqn{kfst} 
\eea
where $B\dn f$,   
a constant\footnote{\label{fnt1} Also called the critical 
magnetic field strength in the literature,
$B\dn f \deq \sqrt{\dsf{2 \me\uf\, c\uv}{\e\sz\, h\cu}}
 \approx \scn{3.6}{16} \, T$ in SI units.}, has the dimension of 
 magnetic field strength to ensure that the EFST $F_{\mu\nu}$ has the required 
 dimension of 
magnetic field strength as well. 
Since $K\sps i$ is a Killing vector field,  $K\sps i\dosw \m \n \deq - K\sps i\dosw \n \m$,
and thus  
\bea
F\dmn\lrs{K\sps i}  \aea 2 \, \z\, \colb{K\sps i\dosw\n\m} \leqn{efst}
\eea  
Papapetrou \cite{papa} showed that if $K$ is a Killing vector field then
$F\dmn\lrs K$ satisfies the  
Maxwell's equations in vacuum.   
See Appendix \ref{app1}.  

Since every linear combination of 
$K\sps 1, K\sps 2$ and $K\sps 3$, shown in \ct{kvs},  is also a Killing vector, 
we consider the symmetric linear combination of $K\sps 1, 
K\sps 2$ and $K\sps 3$, namely
\beas
K(\b) \aea \brf\b{\ldf} (K\sps 1 + K\sps 2 + K\sps 3) \deq \brf\b{\ldf}(0,z-y, x-z,y-x)  
\eeas
where $\b$ is a free parameter, that will be constrained later. 
A straightforward calculation, using the nonzero Christoffel symbols of
flat FLRW \colb{metric}\footnote{\bea
\cfl 0 i i \aea a\, \dot a/c, \quad \cfl i 0 i \deq \cfl i i 0 \deq H/c, \quad i \deq 1, 2, 3, \quad H \dfn \dsf{\dot a}a \quad \leqn{cfs}
\eea}  
shows that the nonzero components of the EFST $F\dmn\lrs{K(\b)}$ are
\bea 
F\dn{0i}\lrs{K(\b)} \aea -F\dn{i0}\lrs{K(\b)}\deq \colb{-}\brf{2\z \,    a\sq \, H}{c} K\tu i(\b), \quad i = 1, 2, 3, \quad H\dfn \dsf{\dot a}a\nn
F\dn{12} \aea  -F\dn{21} \deq F\dn{23} \deq -F\dn{32}   \deq F\dn{31} \deq -F\dn{13}  \deq \colb{-}2 B\dn f \, \b\, a\sq \leqn{efk} 
\eea

The  standard observer\footnote{Also called   comoving observer, \colb{a standard observer} is actually a 
family of subobservers---one observer at each 
 triple of  comoving
spatial coordinates---with the comoving spatial coordinates
of each subobserver being independent of the ``bookkeeper'' 
coordinate $t$.} in flat FLRW spacetime,
has 4-velocity $u\upm \deq (1 ,0,0,0)$ in comoving coordinates.  The  magnetic field 4-vector seen by the observer  is given by 
\cite{subr,tsag,bmts}
\bea
B\upm\uu{(K)} \aea \f12 \dsf{\cla\tu{\m\n\a\b}}{\sqrt{-g}} u\dnn F\dab \deq   
\dsf{2 \, \b \, B\dn f}{ a(t)}\, (0,1,1,1)\leqn{kb}
\eea
where $\cla\tu{\m\n\a\b}$ is the totally antisymmetric 
symbol with $\cla\tu{0123} \deq  1$.  The
subscript $\uu{(K)}$  
denotes that the field  arises out of the Killing symmetry of the spacetime.
Since $B\upm\uu{(K)}$, a 4-vector,
is nonzero in comoving coordinates it is nonzero for every observer and hence 
every
observer sees a nonzero magnetic field arising out of the Killing vectors in the
flat FLRW spacetime.

The projection tensor 
\bea
h\dmn \ada g\dmn - u\dnm u\dnn \leqn{prt}
\eea
can be used to project any tensor onto the spatial hypersurface $S$,
whose tangent plane, at every point on $S$,  is orthogonal to a 
given unit timelike (congruence) 
vector field $u\upm$. 

Projecting $B\upm\uu{(K)}$ onto $\cls$, the spatial hypersurface orthogonal
to the 4-velocity of the standard observer, we obtain $B\upm\uu{(p)}$
\beas
B\upm\uu{(p)} \aea h\upd\m\n B\upn\uu{(K)} \deq \lrr{\de\upd\m\n - u\upm u\dnn} B\upn\uu{(K)} \deq \colb{B \upm \uu{(K)}}
\eeas
confirming the obvious fact that $B\upm\uu{(K)}$ lies in $\cls$.  Thus, in
 comoving coordinates, the standard observer sees a homogeneous magnetic field
\bea
\vec B\uu{(K)} \aea  \dsf{2 \, \b \, B\dn f}{ a(t)}\, (1,1,1) \leqn{kmf}
\eea
that is confined to the spatial hypersurface orthogonal to his world line.    
The magnetic field  in \ct{kmf}  decreases with $a(t)$ in comoving coordinates.  
 Assuming that the observations of   
Pomakov et al \cite[Table 6]{poma} were made in comoving coordinates,
\ct{kmf} is consistent with their results,  which show an increase in 
the strength of the magnetic field with redshift in the primordial 
magnetogenesis model.
The  invariant scalar of the magnetic field in \ct{kmf} is 
 $B\upm\uu{(K)} (B\uu{(K)})\dnm \deq -12\b\sq B\dn f\sq$, which is 
 nonzero  if $\b\neq 0$. 

The  magnetic field $B\upm\uu{(K)}$---generated by  Killing symmetries---is of primordial origin, coming into existence at the birth of the flat FLRW spacetime itself.
As mentioned in Section \ref{sec:cmf},   
Thorne conjectures that the CMF ought to have emerged at big bang, but
does not provide a mechanism by which a magnetic field arises at big bang. The above discussion strengthens   
Thorne's 
conjecture by showing 
that the birth of the CMF at big bang
does not have to be postulated, but can be derived 
as a  consequence of the spacetime's Killing symmetries\footnote{Assuming that the flat 
FLRW spacetime emerged at big bang.}.  

The  electric field 4-vector seen by the standard observer is given by \cite{subr,tsag,bmts}
\bea
E\upm\uu{(K)} \aea F\umn u\dnn \deq \colb{-}\dsf{2 B\dn f \b   H}c (0, z-y, x-z, y-x)
\leqn{elef}
\eea
 
Projecting $E\upm\uu{(K)}$ to $\cls$ we obtain $E\upm\dn{(p)}$
\bea
E\upm\dn{(p)} \aea h\upd \m \n E\upn\uu{(K)} 
\deq (\de\upd\m\n - u\upm u\dnn) E\upn \deq E\upm\uu{(K)} \leqn{elf}
\eea
again confirming the obvious fact that $E\upm\uu{(K)}$ lies in $\cls$.  
Similarly, the projection of the 
covariant derivative of the electric field $\nabla\dnm \lrr{E\uu{(K)}}\dnn $ 
onto $\cls$ is
\beas
\nabla\dnm\sps p \lrr{E\uu{(K)}}\dnn \aea h\dnu  \m\a h\dnu \n\b \nabla\dna \lrr{E\uu{(K)}}\dnb
\eeas

Since the electric field resides in the spatial hypersurface orthogonal to 
the 4-velocity of the standard observer,   
we can covariantly construct the `curl'\footnote{Strictly speaking,  the curl of a 4-dimensional vector field 
is not a vector, but a bivector.  However, both the electric field $E\sbs K$ 
and $\om\upm$, defined in \ct{doo}, lie  in the 3-dimensional hypersurface $\cls$, in the sense that the projection 
of $E\sbs K$ (resp. $\om\upm$) to $\cls$ coincides with $E\sbs K$ 
(resp. $\om\upm$).   
Therefore, 
with some abuse of notation we 
refer to the 4-vector $\om\upm$ as `curl' with the understanding
that we ignore the zeroth component of $\om\upm$.} of the 
electric field in $\cls$, denoted $\om\upm$, as \cite{sub2}
\bea
\om\upm  
\ada   \brf{\cla\tu{\m\n\a\b}}{\sqrt{-g}} u\dnn  \nabla\dna\sps p
\lrr{E\uu{(K)}}\dnb  \deq    
\brf{\cla\tu{\m\n\a\b}}{\sqrt{-g}} u\dnn h\dnu\a\rho h\dnu\b\sg  \nabla\dnr 
\lrr{E\uu{(K)}}\dns   
\leqn{doo}
\eea
where $A\tu{\m\n\a\b}$ is the totally antisymmetric symbol with $\cla\tu{0123} \deq 1$.   It is easy to see that in a locally inertial frame,  $\om\upm \deq (0, \nabla \times \vec E\uu{(K)})$; that is, $\omega$ effectively reduces to the 3-dimensional curl. 

Using \ct{elef} and \ct{doo} we calculate $\omega\upm$ to obtain 
\bea  
\omega\upm  
\aea \colb{-}\brf{4B\dn f\b  }{  c}\bsf{H}{a(t)}(0,1,1,1) \leqn{curl}
\eea
Since the 4-vector 
$\om\upm\neq 0$ in comoving coordinates, every observer sees a nonzero $\om\upm$.  In particular,
the standard observer sees a homogeneous  nonzero curl of the 
electric field \ct{elef} at every point on the spatial hypersurface $\cls$.    

Like the Killing magnetic field $B\upm\sbs K$, the rotational electric field $E\uu{(K)}$---generated by   Killing symmetries---is also of primordial origin, coming into existence at the birth of the flat FLRW spacetime.   
The  
nonzero curl of the primordial  $E\sbs K$ would have induced pervasive vorticity in the charged matter that existed in the 
pre-recombination early universe,   
raising the following speculative question: {\it does the vorticity that the rotational  electric field $E\sbs K$ induces  in  charged matter in the early universe  
provide a viable 
mechanism for seeding the ``axial rotation'' that we observe today 
in ``all successive degrees of accumulation of matter, such as planets,
stars and galaxies''} \cite{gamo}?  See Section \ref{room} for a discussion
of the background for the above question.   

Secondly,  
$\om\upm$, the curl of the $E\sbs K$, is still nonzero today.  Hence, it is reasonable to ask the following question: {\it 
can we detect   
the Killing electric field $E\sbs K$
in the laboratory today?  
}  \colb{In \S \ref{falsifiability} we argue that the answer to the
preceding question may be within current experimental reach.}
The answer to
 the  question depends on the magnitude of the  parameter 
 $\b$ for which we derive bounds in the following discussion.  

The homogeneous magnetic field $B\sbs K$, seen by the standard observer   
is at odds with the underlying assumption of  
isotropy of the flat FLRW spacetime.  
Therefore,  the above derivations  
would be self-consistent only if the Killing electromagnetic field has a negligible
back reaction on the flat FLRW metric.   The self-consistency constraint enables  
us to derive an upper bound
on the strength of the Killing magnetic field, or in other words
an upper bound on   
$\b$. 

The Hilbert stress-energy tensor for the 
electromagnetic field strength tensor arising from 
the Killing field $K(\b)$, shown in  \ct{efk}, is\colb{,  
in} comoving coordinates $(t,x,y,z)$, 
\bea
\colb{\lrr{T\sbs K}}\dmn  (t,0,0,0) \aea     
\ltd \dsf 1 {2\m\sz} \lrs{-\fdn\m\a \fdu\n\a + \f14 \gd\m\n  F\dab  F\uab}\right|\dn{(t,0,0,0)}    \nn
\aea
\dsf{ \b\sq\, B\dn f\sq }{ \m\sz} \mtrx{cccc}{
3   \j 0\j  0 \j 0\\
0\j    \colb{a\sq}  \j  -2a\sq\j -2a\sq\\
0\j  -2a\sq\j   \colb{a\sq}  \j -2a\sq\\
0\j -2a\sq \j -2a\sq \j  \colb{a\sq}} \quad \leqn{tmn}
\eea
\hs{-0.07}On the other hand the nonzero components of the
Einstein tensor for the flat FLRW
metric, in comoving coordinates $(t,x,y,z)$,  are
\def\rez{1 \j 0 \j 0\j 0}  
\def\reo{0 \j 1 \j 0\j 0}  
\def\rew{0 \j 0 \j 1\j 0}  
\def\rer{0 \j 0 \j 0\j 1}  
\bea  
G\dn{00} \aea \dsf{3H\sq}{c\sq}, \quad 
G\dn{11} \deq G\dn{22}\deq G\dn{33} \deq \dsf{2a\ddot a + \dot a\sq}{c\sq}
\leqn{et}
\eea 
We assume that the universe is populated by matter whose energy-momentum tensor, $T\umn\sbs m$,  is proportional to the Einstein
tensor, shown in \ct{et}, and satisfies the Einstein's equations in the
absence of the KEF.  Specifically, 
\bea
T\umn\sbs m \aea \brf{c\tu 4}{8\pi G} G\umn
\eea
The  $T\umn \sbs K$ arising
from the KEF is regarded as a perturbation of $T\umn\sbs m$.  Then
\beas
T\umn\dn{(total)} \aea T\umn\sbs m + T\umn \sbs K  
\eeas  
$T\umn\sbs K$ would be negligibly smaller
than  
$T\umn\sbs m$ and hence  
$T\umn\sbs{total} \approx T\umn\sbs m$
if 
\beas  
\b\sq \j \ll \j  
\bsf{\m\sz c\sq  }{8 \pi G B\dn f\sq} 
\colb{(\ddot a /a + H\sq/2)}
\eeas
We note that the value of the deceleration parameter today is \cite{cama}
\beas
q\sz\aea - \dsf{\ddot a}{a H\sq} \deq - 0.55 
\eeas
Hence the bound on $\b$, derived from self-consistency requirement,
is 
\bea
\b \j \ll \j  
\sqrt{\dsf{ \m\sz c\sq  H\sq}{8\pi G B\uu f\sq}} \ \dfn \ \b\sz  \deq \scn{4.98}{-25}
\leqn{ubob}
\eea
The upper bound on the right in \ct{ubob} is denoted $\b\sz$ for 
later convenience.
Taking the current value of $a\deq 1$, the magnetic field arising out of the Killing vector $K(\b)$, given by
\ct{kmf}, is thus constrained by self-consistency condition to satisfy\footnote{In accordance with the standard terminology,  the symbol $G$   has been used to
denote both the gravitational constant as well as Gauss, the unit of magnetic
field strength.  The discerning reader will determine what the symbol 
stands for from the context. }
\bea
 |\vec B | \deq 2\sqrt 3 \, \b\, B\dn f  \j \ll \j 2\sqrt 3\, \b\sz \, B\dn f\deq 2\sqrt 3 \colb{\bsf{ \m\sz \, c\sq \, H\sq}{8 \pi G}}\tu{1/2} \sim\, 
\scn{6.21}{-4}\, G \qquad \leqn{ubm}
\eea
which is consistent with the 
bounds 
on CMF, mentioned in Section \ref{sec:cmf}. In particular, if the 
Killing magnetic field, shown in \ct{kmf}, is the CMF, denoted $B\dn{CMF}$, then the 
the recent upper bound of $\sim 10^{-10}$ G on $B\dn{CMF}$, derived from the ultrahigh-energy cosmic ray hotspot in the Perseus-Pisces region \cite{nska}, yields the following upper bound on $\b$,
\bea 
2\sqrt 3 \, \b \,  B\dn f  \sim \lrb{B\dn{CMF}} \j \lesssim \j \potm{10}\, G
\quad \implies \quad 
\b\,\lesssim \scn 8 {-32} 
\, \sim \potm 7\, \b\sz \leqn{ubb}
\eea 
which is not 
inconsistent with the self-consistency condition $\b \ll \b\sz$, 
shown in \ct{ubob}.    

Using the lower bounds on the CMF, $B\uu{CMF}$ deduced by Neronov, Vovk \cite{nevo}, Tavecchio et al., \cite{tave} and Dolag et al., \cite{dola}, we can also obtain a lower bound on $\b$, under the assumption that the Killing magnetic field of the flat FLRW spacetime is the CMF
we observe today.  
\bea 
 \lrb{2\sqrt 3 \, \b B\dn f} \sim \lrb{B\dn{CMF}}  \geq \scn 3 {-16} \, G
\quad \mbox{or} \quad 
 \b \j \gtrsim \j  
\scn{2.4}{-37}\leqn{lbb}
\eea
From \ct{ubb} and \ct{lbb}, under the assumption that the 
Killing magnetic field of the flat FLRW spacetime is the CMF, we conclude that 
\bea
-36.61 \j \lesssim\hs{0.1} \log\dn{10}\lrr{\b}\hs{0.1}\lesssim \j -31.10
\leqn{b-bound}
\eea 

The backreaction of the KEF on the flat FLRW metric was not considered
in the above analysis, in which, for self-consistency, the KEF was constrained
to make a negligibly small contribution to the overall stress-energy tensor.  
A more rigorous derivation would involve extremizing the following action,  
\beass
S \aea \int \dfx  \sqrt{-g} \left[\brf {c\cu}{16\pi G}   
\clr \rtd\nn 
\aba - \ltd
\colb{\dsf{1}{4\m\sz}}
\lrc{  
F\umn[\sum_a \colb{ \b\sps a }K\sps a(g\dmn,
\metd \m\n\a)]
F\dmn[\sum_a \colb{ \b\sps a } K\sps a(g\dmn, \metd\m\n\a )]} + \cll\dn m\right] 
\eeass  
\vs{-0.13}
\bea
\leqn{glag} 
\eea

\noindent  
The metric $g\dmn$ that appears above depends on both the electromagnetic fields
arising from the Killing symmetries of $g\dmn$,  as well as the distribution
of other matter in the universe, described by $\cll\dn m$.   The Killing vectors, 
in turn, are determined
by the metric and its derivatives.   The superscript on the Killing vectors 
$K\sps a$ label the different Killing vectors determined by
the metric $\gd \m\n$ and its derivatives $\metd \m\n\a$. 
\colb{Each Killing vector    $K\sps a$ is 
scaled by a dimensionless constant $\b\sps a$; the values of the
$\b\sps a$s are determined by minimizing the action $\cls$.} 
Since \colb{$\b\sps a$s} determines the \colb{strengths of the KEFs}, the metric $\gd\m\n$ depends
on \colb{$\b\sps a$s} as well.  

Minimizing the above action to simultaneously derive both the metric of 
spacetime as well as the KEFs of that metric is a formidable problem. In 
particular, the KEFs arising from the flat FLRW metric alter the overall
stress-energy tensor---introducing off-diagonal terms as shown in \ct{tmn}---and
as a result the assumption of flat FLRW background metric that gave rise
to the KEFs may no longer be valid.  In our analysis, we assumed that the 
spacetime continues to be described approximately by the flat FLRW 
metric---and hence the Killing vectors we consider remain valid Killing 
vectors---if the contributions of the KEFs are negligibly small.   
 
\colb{If the KEF  postulate is indeed valid, then the} 
emergence of a magnetic field and an electric
field with nonzero curl---both of which introduce a preferred direction---shows
that a purely flat FLRW spacetime is untenable.   
The CMF with large correlation
lengths, which have been observed in the universe,  
demonstrates a 
weak deviation from the  
isotropic flat FLRW spacetime,   
at least at the correlation length scales
of the CMF. 
 
The degeneracy in the 
choice of the  basis vectors in the spacelike hypersurface \cls\,  implies degeneracy of
the  Killing magnetic fields that can arise in the flat FLRW spacetime,
with the degeneracy possibly broken either by metric fluctuations
or by the stochastic deviations from isotropy of  
the other electrically charged matter and dipoles in the universe.
\s{{Killing Electromagnetic Field of Schwarzschild Spacetime}}\label{schwarz} 
\noindent \colb{As  is well known \cite{wald},} 
the KEF, $F\dmn[K]$, generated by the timelike  
Killing vector, $K$, of the Kerr metric is proportional to  the 
electromagnetic field strength tensor $F\sps{KN}\dmn$ 
of the Kerr-Newman metric. 
If \colb{the angular momentum of the rotating mass} $J\ra 0$, then the Kerr (resp. Kerr-Newman) metric reduces to
the Schwarzschild (resp. Reissner-Nordstr$\ddot{\mbox{o}}$m) metric. Thus, it is 
also well known that the  Killing electromagnetic field, generated
by  the timelike Killing vector of the Schwarzschild metric, is proportional to the electromagnetic field of the Reissner-Nordstr$\ddot{\mbox{o}}$m metric.

In the following discussion we consider the following converse question about uniqueness: {\it does the back reaction of the KEF of the timelike Killing vector of  the Schwarzschild spacetime  change the Schwarzschild spacetime to a Reissner-Nordstr$\ddot{\mbox{o}}$m spacetime?}   
We show that,  far away from the 
Schwarzschild mass distribution, the back reaction of a sufficiently
weak KEF does indeed change the Schwarzschild metric to Reissner-Nordstr$\ddot{\mbox{o}}$m metric.

Since the Schwarzschild metric of a mass distribution with mass $M$,
\beasm
ds\sq \aea c\sq dt\sq \chi(r) - 
 \dsf{dr\sq} {\chi(r)}  - r\sq \lrr{d\th\sq + \sin\sq(\th) d\vp\sq}, \qquad
 \chi(r) \dfn \lrr{1-\dsf{2GM}{ c\sq r}},  \qquad \leqn{scm}
\eeasm
does not depend on $t$, \colb{$(K\sps t)\upm(\b) \deq \b (1,0,0,0)$} is a Killing
vector of the spacetime, where $\b$ is a dimensionless parameter. 
\newcommand\deab[2]{\dedd a {#1} \dedd b {#2}} 
The EFST of $K\sps t(\b)$ is 
\bea
F\dmn\lrs{\colb{K\sps t(\b)}} \aea 2 \,\z\, \colb{K\sps t \dosw \n\m},  
\qquad
\z \dfn  \dsf{e\colb{c}}{4  \pi\e\sz G M}, \leqn{zta} 
\eea

For brevity, in the rest of the section we omit the dependence on $K\sps t$
and $\b$ and write just $F\dmn$ or $F\umn$.  
The nonzero components of the EFST are 
\bea
F\dn{01}  \aea - F\dn{10}  \deq  
\b  \brf \k {r\sq}, \quad \k \dfn \colb{- } \bsf{\colb{2}\, \z\, G\, M}{c\sq}
\leqn{fk}
\eea 
The nonzero components of $F\umn \deq \gu\m\a\gu\n\b F\dab$ are
\bea
F\tu{01} \aea - F\tu{10} \deq  - \b\brf\k{r\sq} \leqn{ref}
\eea 
We note that $F\umn$ represents a radial electric field, and 
does not contain a magnetic field. 
Both the Schwarzschild metric as well as the KEF \ct{fk}
are {\it static} and 
{\it spherically symmetric}, suggesting
 the following ansatz for the modified metric:  
\bea
ds\sq \aea  \eto{p(r)} c\sq dt\sq   - \eto{q(r)} dr\sq   - r\sq (d\th\sq + \ssth d\vp\sq)
\leqn{mm}
\eea
with the constraint\footnote{Note that $p(r)$ and $q(r)$   
depend on $\b$, although the dependence is not explicitly shown.} that as $\b\ra 0$, we recover the Schwarzschild metric
\bea
\lt \b 0 \eto{p(r)} \aea  \chi(r), \qquad \lt \b 0 \eto{q(r)} \deq \chi\inv(r)
\leqn{cnstrnt}
\eea 
Secondly, the EFST, shown in \ct{fk}, vanishes at spatial infinity.  Therefore, when the 
KEF is turned on in Schwarschild spacetime, the modified metric---as a result of the 
back reaction due to the KEF---will remain asymptotically flat  and   regular at 
spatial infinity  even for nonzero $\b$.    Thus, outside the Schwarzschild radius, we can expand 
$\eto{p(r)}$ as 
\bea
\eto{p(r)} \aea \chi(r) + \b\lrs{\dsf{a\sk}{r^k} + \dsf{a\dn{k+1}}{r\tu{k+1}} + \ldots}\  
\qquad k\geq 1\leqn{psb}  
\eea
where $a\sk$ is the first nonzero coefficient in the series\footnote{
We don't need to include terms of $O(\b\sq)$ and higher in this expansion since they also have the same form as the $O(\b)$ term,
and hence can be merged with the $O(\b)$ term the effect of which would be to replace the  
$\b$ in the $O(\b)$ term with some function $f(\b)$ which can then be redefined as 
our new small parameter $\b' = f(\b)$.}.

Since the metric in \ct{mm} is also independent of $t$, $Q\upm = 
 \b (1,0,0,0)$ is a 
Killing vector of the metric.  The KEF associated with $Q\upm$ is 
\bea 
F\dmn[Q] \aea \colb{2\, \z\, Q\dosw\n\m} \leqn{fq}
\eea
The nonzero components of $F\dmn[Q]$ are 
\bea 
F\dn{01} \aea - F\dn{10} \deq \colb{-}\b\,\z\, \eto{p(r)} \, p'(r)\leqn{fqq}
\eea
Differentiating \ct{psb} we obtain
\bea
 \eto{p(r)} \, p'(r) \aea \brf{2GM}{c\sq}\dsf 1{r\sq}- \dsf{\b}{r\tu{k+1}} \lrs{ k\, a\sk  + (k+1) \brf{a\dn{k+1}}{r}  + \ldots} \nn
 \aea  \brf{2GM}{c\sq}\dsf 1{r\sq}- \dsf{\b}{r\tu{k+1}} \lrs{ k\, a\sk  + h(r)}, 
 \qquad k \geq 1 \leqn{fdn}
 \eea
where
 \beas
 h(r) \ada (k+1) \brf{a\dn{k+1}}r + \ldots
 \eeas
 Since $h(r) \ra 0$ as $r\ra \infty$, there exists a $r\sz$ such
 that for $r> r\sz$, $\lrb{h(r)} < \lrb{a\sk}.$  Then, for $r>r\sz$
\beas
\lrb{k \, a\sk + h(r)} \alea k \lrb{a\sk} + \lrb{h(r)} \ \leq \ (k+1)\lrb{a\sk}
\eeas
We consider the weak-field regime in which, for $r > r\sz$, 
\beas
\lrb{\dsf\b{r\tu{k+1}} (k\, a\sk + h(r))} \alea 
\dsf\b{r\sz\tu{k+1}} (k+1)\lrb{a\sk} \ \ll \ \brf{2GM}{c\sq} \dsf 1 {r\sq} 
\ \leq \brf{2GM}{c\sq} \dsf 1 {r\sz\sq} 
\eeas
or in other words
\bea
\b \j \ll \j \dsf{2GM}{c\sq (k+1) \lrb{a\sk}} r\sz\tu{k-1}, \qquad 
k\geq 1 \leqn{wfl}
\eea
In the weak-field regime, and for $r > r\sz$,  the $O(\b)$ term in 
\ct{fdn} is negligibly smaller than the leading term and hence can be omitted.  The
nonzero components of the  EFST
 \ct{fq},  are then
\bea
F\dn{01} \aea - F\dn{10}  
\deq \colb{-}\b \brf A{r\sq}, 
\qquad    
A \dfn  r\dn s \z, \quad r\dn s \dfn \dsf{2GM}{c\sq}  
\leqn{wff} 
\eea

Two of the nonzero components of the Ricci tensor of the modified metric \ct{mm} are 
\beasm
R\dn{00} \aea \dsf{\eto{p-q}}2 \lrs{p'' + \dsf{(p')\sq}2 + 2\brf{p'}r - \dsf{p'q'}2}, \quad 
R\dn{11} \deq -\dsf{p''}2 - \dsf{(p')\sq}4 + \dsf{p'q'}4 + \dsf{q'}r
\quad   \leqn{f2cor} 
\eeasm
where $p'$ and $q'$ denote derivatives of $p$ and $q$ with respect to $r$.

The corresponding components of the stress energy tensor of  the EFST \ct{fq}  
\bea
T\dmn \aea \dsf 1{2\m\sz} \lrs{-F\dn{\m\a}F\dnu\n\a + \f14 \gd\m\n
F\uab F\dab} \leqn{set}
\eea
are 
\bea
\ba{rclrcl} 
T\dn{00} \aea  
\eta \eto{2p-q}\colb{,} \hs{0.5} \j 
T\dn{11} \aea - 
\eta \eto{p}, \hs{0.5} \eta \dfn \ds\dsf{(\b\z p')\sq}{4\m\sz} 
\ea \leqn{f2cos}  
\eea
From \ct{f2cos} we note that 
\bea
\dsf{T\dn{00}}{\eto{p(r)}} +\dsf{T\dn{11}}{\eto{q(r)}} \aea 0
\leqn{ppqe}  
\eea
Outside the compact spherical mass $M$ that gives rise to the Schwarzschild metric \ct{scm}, Einstein's equation is 
\bea
R\dmn - \f12 \gd\m\n R \deq \brf{8\pi G}{c\tu 4} T\dmn \leqn{ee}
\eea
where $R$ is the Ricci scalar and $T\dmn$ is the stress energy tensor of the KEF shown in  \ct{set}.   Since $T\dmn$ is traceless\footnote{\colb{
$T\upm{}\dnm \deq - F\dmn F\umn + \dsf 1 4 \deud \m\m F\dab F\uab \deq 0$}} we conclude that 
the Ricci scalar vanishes\footnote{\colb{Taking the trace on both sides of \ct{ee} 
and noting that $T\upm{}\dnm \deq 0$, we obtain $-3R \deq \brf{8\pi G}{c\tf} T\upm{}\dnm \deq 0$}} and Einstein's equation becomes
\bea
R\dmn \aea \brf{8\pi G}{c\tu 4} T\dmn \leqn{eee}
\eea
From \ct{ppqe} and \ct{eee} we conclude that 
\beas
\dsf{R\dn{00}}{\eto{p(r)}} +\dsf{R\dn{11}}{\eto{q(r)}} \aea 0
\eeas
or using \ct{f2cor} we obtain the following equation at all finite $r$
greater than the radius of the spherical mass $M$,
\beas
p'(r) + q'(r)) \aea 0
\eeas
That is, $p(r)+q(r)$ is independent of $r$.    
We are trying to compute the modified metric when the EFST, shown in 
\ct{fk}, is turned on in Schwarzschild spacetime.  Since the EFST vanishes
as $1/r\sq$ as $r\ra \infty$,  
the spacetime inhabited
by the KEF and the compact spherical mass $M$ is asymptotically flat or 
\beas
\lt r \infty p(r) \aea \lt r\infty q(r) \deq 0
\eeas
from which we conclude 
\bea
p(r) + q(r) \equiv 0 \implies p(r) \deq - q(r) \leqn{ppqiz}  
\eea
Using \ct{ppqiz} the remaining nonzero components of the Ricci 
tensor   are 
\bea
R\dn{22} \aea 1 -  \lrr{r\, \eto{p(r)}}' \qquad R\dn{33} \deq \ssth\lrs{1 -  \lrr{r\, \eto{p(r)}}'}\leqn{rcrt}  
\eea  
In the weak field regime and for $r> r\sz$,  
$T\dn{22}$ is given by
\bea
T\dn{22} \aea   
\dsf{\k}{ r\sq} 
\qquad \k \dfn \b\sq\brf{\z\sq\, r\dn s\sq}{4\,\m\sz} 
\leqn{rcset}  
\eea
Using \ct{rcrt} and \ct{rcset} 
the Einstein equation $R\dn{22} \deq \brf{8\pi G}{c\tu 4} T\dn{22}$ becomes
\bea
1- \lrr{r\, \eto{p(r)}}' \deq \dsf \xi {r\sq}, \qquad \xi\dfn 
\b\sq\brf{8\pi G}{c\tf} 
\dsf{(\z r\dn s)\sq}{4 \m\sz  }  \leqn{fe}
\eea
The solution of \ct{fe} is 
\beas
\eto{p(r)} \aea 1 + \dsf C r + \dsf{\xi}{r\sq}
\eeas
where $C$ is an arbitrary constant. The constraint \ct{cnstrnt} implies
that 
\beas
C \deq - \dsf{2GM}{c\sq}
\eeas
Therefore, we conclude that in the weak field regime and for $r>r\sz$,
 the modified metric \ct{mm} has the form
\beasm
ds\sq \aea \lrr{1- \dsf{2GM}{c\sq r} + \dsf \xi{r\sq}} c\sq dt\sq - 
\dsf{dr\sq} {\lrr{1- \dsf{2GM}{c\sq r} + \dsf \xi{r\sq}}} - r\sq \lrr{d\th\sq +
r\sq\ssth d\vp\sq}\nn
\leqn{mmf}
\eeasm
which is the Reissner-Nordstr$\ddot{\mbox{o}}$m metric, with mass $M$ and charge $Q$
\fs{9}{10}{\bea 
ds\sq \aea \lrr{1- \dsf{2GM}{c\sq r} + \dsf {GQ\sq}{4\pi\e\sz c\tu 4\, r\sq}} c\sq dt\sq - 
\dsf{dr\sq} {\lrr{1- \dsf{2GM}{c\sq r} + \dsf {GQ\sq}{4\pi\e\sz c\tu 4\, r\sq}}} \nn
\aba - r\sq \lrr{d\th\sq +
r\sq\ssth d\vp\sq}\nn
\leqn{rnm}
\eea}  
where 
\bea
Q \aea \b  \bsf{2\sqrt 2\, \pi \,\z\, r\dn s}{\m\sz c} 
\leqn{chrg}
\eea
In particular, in the weak field regime an observer,  far from a Schwarzschild black hole ($r>r\sz$), 
would conclude that it is actually a Reissner-Nordstr$\ddot{\mbox{o}}$m black hole containing
electric charge given by \ct{chrg}.  {We note that the above derivation
remains valid for any spherical distribution of stationary matter.}
Thus every spherical distribution of stationary matter---regardless of the amount of 
matter---appears electrically charged to a distant observer in the weak 
field regime.   Finally, we note that we restricted attention to the weak field 
regime only to make the calculation more tractable.  It is worth exploring if
we can arrive at the same conclusion even when the weak-field assumption is   dropped.  

Penrose's {\it weak cosmic censorship hypothesis} yields an upper bound on 
$\b$, even if it is  
a rather weak bound.  
In the Reissner-Nordstr$\ddot{\mbox{o}}$m metric \ct{rnm}, if 
\bea
\dsf{Q }{M} \aga \sqrt{4\pi\e\sz G} \leqn{wcch}
\eea
then $\gd00 > 0$ and  the spacetime has no horizon.  And yet, $r=0$ is a singularity---the so-called {\it naked singularity} since 
it  \colb{would not be }
hidden inside a horizon\colb{, when \ct{wcch} holds}.  Assuming the validity of  Penrose's  weak  cosmic censorship hypothesis \cite{penr}, which posits that naked singularities do not exist, we obtain the condition 
\bea
\dsf Q M \alea \sqrt{4\pi\e\sz \, G}  \leqn{wcch}
\eea
From \ct{chrg} and \ct{wcch} we obtain the following upper bound on 
$\b$,
\bea
\b \alea  \colb{\sqrt{4\pi\e\sz G}\bsf{\m\sz \, c\, M}{2\sqrt 2 \,\pi\,\z\, \rs}} 
\leqn{cchb}
\eea
\s{\colb{Discussion}}\label{concl}
\noindent\colb{
We begin the section by verifying the gauge invariance
of  KEF, in \S \ref{gauge}.  We then discuss the plausibility of the KEF and
KEC postulates,  in \S \ref{plausibility}.
We  also argue
that  falsifying the KEF postulate  may be 
 within experimental
reach, in \S \ref{falsifiability}.   In \S \ref{mtfrt},
drawing upon Rainich's result \cite{rain},  
we discuss the etiological question  associated
with the KEF and KEC postulates.  
In \S \ref{cwc}, we discuss the relevance of
  the aforementioned
transmutation of 
the Schwarzschild metric and the KEC postulate, to
Misner and Wheeler's program of realizing ``charge without charge''
\cite{miwh}.  Finally, in \S \ref{mfkv}, we  discuss the features of KEF and KEC that 
distinguish  magnetogenesis
sourced by Killing vectors from previous proposals
for primordial magnetogenesis.}

\ss{\colb{Gauge invariance}}\label{gauge}
\noindent\colb{Under a gauge transformation of the Killing vector $K\upm \ra 
\ti K\upm \deq K\upm + \pl\upm \L$,  $F\dmn \ra \ti F\dmn$,
and
\beasm
\ti F \dmn \aea \z\lrs{\ti K\dosw\n\m - \ti K\dosw\m\n} \deq 
F\dmn + \z \lrc{\lrs{\pmd\pnd \L - \cfl \l \m\n \pl\dnl \L} - 
\m \leftrightarrow \n} \deq F\dmn\nn \leqn{ginv}
\eeasm
Thus, the EFST $F\dmn$ is invariant under a general 
gauge transformation.}   

\colb{One can easily verify that $\ti F\dmn$ also satisfies Maxwell's 
equations. Specifically, 
\bea
\ti F\dn{\a\b;\m} + \ti F\dn{\m\a;\b} + \ti F\dn{\b\m;\a} \aea 0\leqn{hmax}\\
\ti F\umn\dso\n \aea \dsf{J\upm\sbs K}{\e\sz\, c}, \qquad J\upm\sbs K \deq 2\, \e\sz\, c\, \z\,
R\umn K\dnn \leqn{imax}
\eea
Equation \ct{hmax} follows immediately from \ct{ginv} and the proof of \ct{me2}, presented in Appendix \ref{app1}. 
\bea
\ti F\dn{\a\b;\m} + \ti F\dn{\m\a;\b} + \ti F\dn{\b\m;\a} \aea
 F\dn{\a\b;\m} +  F\dn{\m\a;\b} +  F\dn{\b\m;\a} \deq 0 \leqn{mfe}
\eea
Equation \ct{imax} also 
follows immediately from \ct{ginv} and the proof of \ct{me1} presented
in  Appendix \ref{app1}. 
\bea
\ti F\umn\dso\n \aea  F\umn\dso\n \deq \dsf{J\upm\sbs K}{\e\sz\, c} 
\deq 2 \, \z\, R\umn K\dnn\leqn{impeq}
\eea
We note that, in \ct{impeq}, the source term, $2\,\z\, R\umn\, K\dnn$, is written in terms of 
the Killing vector $K\dnn$ and not the transformed vector $\ti K\dnn$,
although the left hand side $\ti F\umn\dso\n$ is written in terms of
the transformed tensor.  We elaborate below.
}

\colb{
Let $\cls$ denote the set
of all twice continuously differentiable scalar fields on the 
spacetime manifold.  Define,
\bea
\cls\sz \aea \lrc{\L\in \cls \ | \ \nk\m\nk\n \L \deq 0,  \ \m,\n = 0, 1, 2, 3}
\leqn{defsz}
\eea
The set of all gauge transforms of a Killing vector $K$ is  
\beas
\clg(K) \aea \lrc{\ti K\upm \, | \, \ti K\upm \deq K\upm + \pmu \L, \ \L \in \cls  }
\eeas
Also, we define
\bea
\clg\sz(K) \aea \lrc{\ti K\upm\, |\, \ti K\upm \deq K\upm + \pmu \L, \ 
\L\in \cls\sz}, \qquad \clg\so(K) \dfn \clg(K)\setminus \clg\sz(K)\nn
\leqn{g0g1}
\eea
First, we note that for a gauge transform $\ti K\upm \deq K\upm + 
\pmu\L\in \clg(K)$, 
\bea
\ti K\dosw\n\m + \ti K\dosw\m\n \aea \lrs{K\dosw\n\m + K\dosw\m\n}
+ 2 \, \nk\m\nk\n \L \deq 2 \, \nk\m\nk\n \L \leqn{kvii}
\eea
From \ct{kvii}  and \ct{defsz} we conclude that every vector field in 
$\clg\sz(K)$ is a Killing vector field.  Conversely, none of the vector fields
in $\clg\so(K)$ is a Killing vector field.  {\it In general, the gauge transform 
of a Killing vector is not necessarily a Killing vector.}
}

\colb{
A gauge transform $\ti K\upm\in \clg\sz(K)$, satisfies the following
important property:
\bea
R\umn \ti K\dnn \aea R\umn K\dnn \leqn{tikk}
\eea
where $R\umn$ is the Ricci tensor.}

\colb{
To prove the above property, using \ct{ginv} and recalling that $\ti K\upm$
is a Killing vector, we note that 
\bea
\ti F\dmn \aea  \z\, (\ti K\dosw\n\m - \ti K\dosw\m\n) \deq 
2  \,\z \, \ti K\dosw \n\m \deq F\dmn \deq 2 \, \z\, K\dosw\n\m, \ 
\implies \ \ti K\dosw\n\m \deq K\dosw\n\m\nn 
\leqn{tikeqk}
\eea
For an arbitrary Killing vector $W\upm$, using the symmetries of the Riemann tensor, we have
\bea
R\umn W\dnn \aea 
\gu\m\a   R\mk\l{\n\l\a} W\upn \deq \gu\m\a \lrr{W\mk\l {;\a;\l}
-W\mk\l {;\l;\a}} \nn
\aea 
\gu\m\a \gu\l\rho \lrr{W\dn{\rho;\a;\l} + W\dn{\l;\rho;\a}} 
\qquad \leqn{forw}
\eea
Using \ct{mfe} and \ct{tikeqk} we have
\bea
\ti K\dn{\rho;\a;\l} + \ti K\dn{\l;\rho;\a } \aea 
- \ti K\dn{\a;\l;\rho}, \qquad
 K\dn{\rho;\a;\l} +  K\dn{\l;\rho;\a } 
 \deq 
 -  K\dn{\a;\l;\rho} \leqn{fork}
\eea
From \ct{forw}, \ct{fork} and \ct{tikeqk} we have
\beas
R\umn \ti K\dnn \aea - \gu\m\a \gu\l\rho \ti K\dn{\a;\l;\rho} 
\deq - \gu\m\a \gu\l\rho  K\dn{\a;\l;\rho} \deq R\umn K\dnn
\eeas}
 \colb{\!\hspace{\parindent}Finally, we return to \ct{impeq}, which states that the gauge
transform $\ti F\dmn$ obeys Maxwell equation, in which the 
current density (right hand side) is expressed in terms of the
untransformed field $K\upm$, and not the transformed field
$\ti K\upm$.  Given a gauge transform $\ti K\upm$, 
we need to specifiy how one can 
construct the source term $2\, \z\, R\umn\, K\dnn$ in 
\ct{impeq}.  We note that $\ti K\upm \deq K\upm + \pmu \L$
for some $\L\in \cls$.  Let $\L'$ be a solution of the following 
equation: 
\bea
\nk\m\nk\n \L \aea \f12\lrs{\ti K\dosw\m\n + \ti K\dosw\n\m} \leqn{getl}
\eea
We note that if $\ti K\upm\in \clg\sz(K)$, then $\L' \deq 0$ is a 
possible solution of \ct{getl}.  For $\ti K\upm\in \clg\so(K)$, $\L' \neq 0$.
Consider
\bea 
Q\upm \aea \ti K\upm - \pmu\L' \deq K\upm + \pmu \L'', \qquad 
\L'' \deq \L - \L', \leqn{forq}
\eea 
Thus, $Q\upm$ is a gauge transform of $K\upm$ and 
from the definition of $\L'$ we know that $Q\upm$ is a 
Killing vector.  Therefore
\beas
Q\dosw\n\m + Q\dosw\m\n \aea 2 \nk\m\nk\n \L'' \deq 0
\eeas
showing that $Q\upm\in \clg\sz(K)$.  Therefore, from \ct{tikk},
we conclude that $R\umn Q\dnn \deq R\umn K\dnn$.  In summary,
we construct the source term $2\,\z\,R\umn K\dnn$ in \ct{impeq},
using $\ti K\upm$ as follows. First, solve 
 \ct{getl}, and then 
using the solution $\L'$  construct $Q\upm$,
as in \ct{forq}.  $R\umn Q\dnn$ yields the required $R\umn K\dnn$. 
}

\ss{\colb{Plausibility}}\label{plausibility}
\noindent\colb{The main plausibility argument for the postulates
is that they can, if valid, potentially  explain two enigmas---the 
CMF and the axial rotation of all matter.   As mentioned in \S\ref{back},
Thorne writes \cite{thor},``All attempts to understand how galactic magnetic
fields could have arisen ... face difficulties, which seem ... insurmountable.
The only way in which the existence of the magnetic fields can be
understood ... is by assuming that they ... were created in the big bang.''
However, Thorne does not explain {\it why} magnetic fields should be  created in the big bang.  Assuming that the flat FLRW
spacetime emerged at---or shortly after---big bang, the KEF postulate, 
if valid, could potentially  explain  why Thorne's conjecture  could be true.  The second argument that makes KEF an appealing candidate for CMF
 is that the primordial KEF does not depend 
on any model of particle  physics in the early universe.  Finally, the CMF is 
ubiquitous and a natural 
source for the CMF should pervade the 
entire intergalactic region, as KEF does.}

\colb{The sourcing mechanism for the pervasive 
 vorticity---observed at all length scales---remains unknown.   
 If the KEF postulate is valid, and the early universe was a 
flat FLRW spacetime, then the rotational KEF in flat FLRW spacetime 
could potentially
provide a natural and ubiquitous sourcing mechanism for vorticity.}

\colb{A curious plausibility argument for the KEC, $J\upm\sbs K$,
stems from the observation that all known
electrically charged particles have nonzero rest mass.  For some, 
as yet unknown reason, 
existence of 
massless electrically charged particles seems to be forbidden.  $J\upm\sbs K$
is nonzero---and gives rise to nonzero charge density---only where
the Ricci tensor is nonzero and hence  the stress-energy tensor is 
nonzero\footnote{\colb{In 4-dimensional spacetime, if Einstein tensor
$G\dmn$ vanishes, then so does the Ricci tensor $R\dmn$.  
\beas
G\dmn \aea  0 \implies R\dmn \deq 
\f12 \gd\m\n R \implies R \deq 2 R \implies R\deq 0 \implies
G\dmn \deq R\dmn
\eeas}
}. Thus, the KEC $J\upm\sbs K$  stipulates that the 
presence of nonzero charge density at a location
is necessarily accompanied by 
non-vanishing gravitational source at the location, 
consistent with the observed nonzero rest mass of all 
electrically charged particles.
}
\ss{\colb{Falsifiability}}\label{falsifiability}
\noindent\colb{If the Killing electric field of the flat FLRW
 spacetime is indeed real, then it accelerates a point charge,
 such as an electron\footnote{\colb{The energy loss due to radiation from
the accelerating electron is   $\sim \pot{-116}$ W/s  
  and will be ignored.}}.  Using \ct{elef}, and taking $\b \sim \pot{-32}$,
 the acceleration $a$ of an electron, induced by Killing electric field, is
 \beas
 a \j\sim\j \dsf{e\, \b  B\dn f H}{\me  c} \sim \pot{-31} \, \mbox{m}\d\mb{s}\tu{-2}
 \eeas
 Over a year's time, the acceleration induces a displacement $d$ that is approximately
 \bea
 d \j \sim \j \pot{-16} \, \mbox{m.} \leqn{dispd}
 \eea 
The state-of-the-art laser interferometers \cite{abbo} are able to detect displacements as 
small as   
 $\pot{-19}$ m (see footnote\footnote{\colb{For dynamic measurements,
 such as in LIGO,  the strain sensitivity 
around 100 Hz is  
less than $\pot{-23} /\sqrt{Hz}$, which, for an 
arm length of 4 km, yields a displacement sensitivity of  $\scn 4 {-20}$ m.}}), 
making it plausible that, with a suitably designed experiment, the measurement of the displacement $d$ is  within current experimental
reach.}

\colb{While it is possible to shield an electron from   electromagnetic
fields that arise from external charges and currents, an electron cannot be shielded from the Killing electric field, which arises out of spacetime itself.  Hence, the 
effect of KEF can likely be isolated.  Assuming the correctness of the bounds in 
\cite{nska,nevo,tave,dola}, non-detection of
the predicted KEF-induced displacement, corresponding to the range \ct{b-bound},
would show that the CMF is not KEF, and would effectively falsify  
the KEF postulate.}

\colb{
 The KEC $J\upm\sbs K \propto R\umn K\dnn$, is a  weak current.  Even in  
extremal objects such as neutron stars, which have matter density of the order
 of $\pot{18}$ kg/m$\cu$ \cite{latt}, $R\umn \sim \pot{-8}$ m$\tu{-2}$; the 
 nontrivial magnetic field and plasma surrounding neutron stars make it 
 challenging to isolate the effect of a weak  KEC.  The 
 other arena in which the stress-energy tensor is significant is the early universe.
 We are currently investigating possible approaches to discern the signature of 
 $J\upm\sbs K$   
 from the available data about the early universe. 
}
\ss{\colb{Maxwell tensor from Ricci tensor}}\label{mtfrt}
\noindent\colb{In 1925, Rainich \cite{rain} derived an important  result,
which provides insight into the deep connection between 
electromagnetism and geometry. 
Specifically, he showed that a spacetime inhabited only by a source-free electromagnetic field is a {\it Rainich-Riemann manifold}---a
Riemannian manifold in which the Ricci tensor is non-null\footnote{\label{fn:rc}
\colb{$R\dab R\uab \neq 0$.}} and satisfies
the Rainich conditions\footnote{\colb{Rainich conditions on the Ricci 
tensor are
\bea
R\mk\m\m \aea 0, \quad R\kk\m\a R\mm\n\a \deq \de\km\m\n
\lrr{\f14 R\drs R\urs}\quad R\kk 00 \geq 0 \leqn{alg}\\
\chi\dnr\dso\sg - \chi\dns\dso\rho \aea 0, \qquad 
\chi\dnb \deq \dsf{\sqrt{-g} \, \cla\dn{\b\l\m\n} \, R\tu{\l\g;\m} \,R\km\g\n }{R\drs R\urs} \leqn{diff}
\eea
Note that the algebraic conditions \ct{alg} as well as the differential
condition \ct{diff}, are purely geometric constraints  that involve only
the Ricci tensor.  As before, $\cla\dn{\b\l\m\n}$ is the antisymmetric
symbol.}}---and further that the 
source-free electromagnetic field 
is uniquely determined
by the Ricci tensor and its derivative, up to an overall global phase (constant
of integration)\footnote{
\colb{Specifically, 
the Maxwell tensor (electromagnetic field strength tensor) can be written in terms of the Ricci tensor as
follows.
\beas
F\dmn \aea \xi\dmn \cos \a + *\xi \dmn \sin\a
\eeas
where $\xi\dmn$ is the extremal ``Maxwell square root''
of the Ricci tensor obtained as the solution of the equation
\beas
\xi\dab \xi\dmn 
\aea   - \f12 E\dn{\a\b\m\n} - \f12 
\dsf{E\dn{\a\b\g\de}E\km{\m\n}{\g\de}}{(R\dn{ab}R\tu{ab})\tuh}, \qquad \mbox{where}\\
E\km{\tau\rho}{\m\n} \ada 
\f12 \lrr{- \de\km\tau\m R\km\sg\n + \de\km\sg\m R\km \tau\n
- \de\km\sg\n R\km\tau\m + \de\km r\n R\km\sg\m}
\eeas
\colb{and $*\xi\dmn \deq \f12 \sqrt{-g}\cla\dn{\m\n\a\b}\xi\uab$ is the
dual of $\xi\dmn$.  $\a$ is also determined  by $R\dmn$, up to a constant of integration, and is the solution
of the equation 
\beas
\dobydo{\a}{x\upb} \aea \dsf{\sqrt{-g}\, \cla\uu{\b\l\m\n} R\tu{\l\g;\m}R\km\g\n}{R\drs R\urs}
\eeas}}.}.  We call a non-null Ricci tensor that satisfies the Rainich
conditions, a {\it Rainich-Ricci tensor}. }

\colb{Rainich's result, hailed as an ``already unified theory'' by Misner and Wheeler \cite{miwh},  rigorously showed that source-free electromagnetism ($F\dmn$) can be expressed entirely in terms of geometry (in terms of $R\dmn$ and its derivative). The result has
a profound implication:  {\it what we perceive physically  
as---interpret as---a source-free
electromagnetic field is, at a more fundamental level, merely    a  geometric structure constructed from the
Ricci tensor and its derivative in  a special
Riemannian
manifold---the Rainich-Riemann manifold.  
}    
Thus, source-free electromagnetism does not arise out of---and is not separate from---geometry.   {\it Source-free electromagnetism is
geometry.}}  

\colb{Rainich's result provides a possible answer to the following etiological
question that confronts the KEF postulate: 
{\it what is the mechanism---the
manner of causation---by which a physical electromagnetic field (KEF) arises  out
of geometry (Killing vector fields)?}   Rainich's result suggests the following
conjecture, as answer to the etiological question: {\it  Killing 
electromagnetic field is the way we physically perceive---the way we interpret---the special geometric structure constructed from the 
 Killing vector fields, namely the covariant derivatives of the Killing
 vectors.
}
KEF does not arise out of geometry. {\it KEF is geometry}. Similarly, we 
conjecture that KEC 
does not arise out of geometry.  {\it KEC is the way we perceive a 
special geometric structure constructed from the second covariant
derivatives of the geometric Killing vector fields}.}

\ss{\colb{``Charge without Charge''}}\label{cwc}
\noindent\colb{Rainich's calculations were followed up by  Misner
 and Wheeler \cite{miwh}, who speculated
 that  the universe has nontrivial topology, and ``wormholes'' in particular.
In a Rainich universe, inhabited only by source-free electromagnetic field, the electric lines of force do not terminate 
at charges; however, an observer without the ability to probe spacetime with
sufficient resolution would interpret the radial lines of force emerging from (or 
going into) the mouth of a sufficiently small wormhole as  lines of force
emerging from (or going into) a point charge. Such an observer---Misner and Wheeler 
argued---would perceive the mouth of a wormhole as 
a point   ``charge'', ``without'' the actual existence of point
``charge''. }

 \colb{The postulated KEC, $J\upm\sbs K$, provides a different mechanism
 for the emergence of charge density $J\uz\sbs K$---from geometry---without 
 requiring the universe to have nontrivial topology (and wormholes).  In other 
 words, the KEC realizes Misner and Wheeler's vision of ``charge without charge''
 even in a universe with a trivial topology.}
 
\colb{ If the KEF postulate is valid, then \ct{ref} is an actual, 
physical electric field, denoted  $E\uu S$.  
$E\uu S$ exerts  electric force on a charged particle making the particle's  motion 
deviate from geodesic trajectory, while an uncharged particle 
moves along  geodesic trajectory---
exactly  as we expect  
in a Reissner-Nordstrom spacetime. Thus, the `charge'
\ct{chrg} 
 not only formally 
resembles the
Reissner-Nordstrom charge but also physically behaves like a real  
 Reissner-Nordstrom charge and, hence, can be regarded as `charge without
charge'.  
  Lacking a good estimate of $\b$, however,
 we are unable to estimate the magnitude of the charge.
 }

\ss{\colb{Magnetogenesis from Killing vectors}}\label{mfkv}
\noindent\colb{Assuming that the two (KEF and KEC) postulates
are indeed valid, the KEF has two features that make it an 
appealing candidate for CMF.  First, the KEF does not depend
on any particle-physics model of the early universe.  In fact, it
does not even rely on the big bang model.  The only underlying
assumption is that
the spacetime is described by the flat FLRW metric.  Secondly,
since the KEF arises everywhere in spacetime, it can provide a
natural explanation for the large coherence
lengths of the CMF. }
\s{Future Work}\label{future} 
\noindent\colb{In this section we describe our ongoing work
as well as a few problems that warrant
further investigation. As we argued in \S \ref{mtfrt}, the  detection of the KEF in flat FLRW may be within current experimental reach.    An 
important follow-up task would be to design  a suitable experiment for
the detection of KEF.  We are currently exploring the possible approaches to discern the signature of KEC 
from the available data about the early universe, and also establish
good upper and lower bounds on the strength of the Killing electric
field in Schwarzschild spacetime. Assuming that the CMF is KEF,
we used 
the current observational bounds on CMF  to derive
upper and lower bounds for KEF in flat FLRW spacetime
(see \ct{b-bound}); however, it 
 would be of greater interest to obtain 
an {\it ab initio} estimate of the strength of the KEF,  by extremizing the action \ct{glag}. 
}

\colb{The primordial rotational electric field, described in \S \ref{fflrw}, generates vorticity.  It is worth investigating if the  conjectured
primordial vorticity generated by Killing vectors can provide
the initial conditions needed in Harrison's magnetogenesis model.   Further, 
the possible role of the conjectured vorticity  generated by Killing vectors 
in structure formation  
also warrants further investigation.}

\colb{
Finally, it would be of interest to study the KEFs and KECs arising from
the Killing vectors that we have not considered---in both the Schwarzschild and
other spacetimes.  As we remarked in Appendix \ref{app1}, it would also be of interest to extend the KEF and KEC postulates to  
 conformal Killing vectors and study the implications of the extended postulates.
}
\colb{\s{Acknowledgment}
\noindent The author is grateful to the referees for the many incisive comments and
questions, which have helped improve both the presentation and content
of the paper.}

\vspace{6pt} 

\appendix
\section[\appendixname~\thesection]{Killing Vectors and Maxwell's Equations}\label{app1}

\noindent Consider the field strength tensor $F\dmn$ derived
from the covariant derivative of a   
Killing vector $K\upm$
\bea
F\dmn \ada \zeta \colb{(K\dn{\n;\m} - K\dn{\m;\n})\deq 2 \, \z\, K\dosw\n\m, }
\eea
where 
$\z$ is a dimensionful constant that endows $F\dmn$ with
the correct dimension; see the discussion in Section \ref{schwarz}.
 The second equality follows from the Killing
 equation
$K\dosw\m\n + K\dosw\n\m\deq 0$.  
We show that $F\umn$  
satisfies  
Maxwell's equations  
\bea
F\umn\,\dn{;\n} \aea \dsf{J\upm}{\e\sz c} \leqn{me1}\\
F\dowsr\a\b\m + F\dowsr \m\a\b + F\dowsr \b\m\a \aea 0\leqn{me2}
\eea
We verify \ct{me2} first.
\bea
\brf 1 \z  \lrs{F\dowsr \a\b\m + F\dowsr \m\a\b + F\dowsr \b\m\a }\aea 
\colb{2K\doswsr  \b\a\m + 2K\doswsr  \a\m\b+ 2K\doswsr \m\b\a}
\leqn{yai}
\eea
Using the Killing equation
$K\dosw\b\m \deq -K\dosw\m\b$   
and noting that 
\bea
K\doswsr \m\a\b - K\doswsr \m\b\a \aea \rmn\l\m\a\b K\dnl
\leqn{rmn1}
\eea 
where $\rmn\l\m\a\b$ is the Riemann tensor,  we 
conclude that
\beas
\brf 1 \z  \lrs{F\dowsr \a\b\m + F\dowsr \m\a\b + F\dowsr \b\m\a} 
\aea (\colb{\rmn\l \b\a\m + \rmn\l  \a\m\b + \rmn\l \m\b\a})K\dnl  
\deq 0
\eeas
The last equality follows from the identity 
\bea
\rmn\l \b\a\m + \rmn\l  \a\m\b + \rmn\l \m\b\a \aea 0 \leqn{rmn2}
\eea

Next  we turn to \ct{me1}.  
Since $\gu\a\b\dn{\ \ \ ;\m} \deq 0$, 
\bea
{F\umn}\dn{;\n} \aea \lrr{\gu\m\a \gu\n\b F\dab}\dn{;\n} \deq 
\gu\m\a \gu\n\b F\dn{\a\b;\n} \leqn{use1}
\eea   
Recalling that $F\dab \deq 2\,\z\, \colb{K\dn{\b;\a}}$, and using \ct{me2}
and \ct{yai} we have
\bea
\gu\n\b F\dn{\a\b;\n}  
\aea 
 2\z R\upd\l\a K\dnl \leqn{use2}
\eea
where $R\dn{\l\a}$ is the Ricci tensor.   
Therefore, using \ct{use1}
and \ct{use2},  and the fact that the Ricci tensor is symmetric, we have
\bea
{F\tu{\m\n}}\dn{;\n} \aea  \dsf{J\upm\colb{\sbs K}}{\e\sz c}, \qquad \mb{ where }
\qquad J\upm\colb{\sbs K} \dfn  2 \e\sz c\, \z\, R\umn K\dnn 
\leqn{maxe} 
\eea
If the Ricci tensor vanishes, as in vacuum, then one obtains the
Maxwell's equations in vacuum.  On the other hand, nonvanishing 
Ricci tensor  gives rise  
to the Killing  current density 4-vector $J\upm\colb{\sbs K}$, shown in \ct{maxe}.  \colb{The subscript $K$ in $J\upm\colb{\sbs K}$ indicates
that the current density emerges from a Killing vector field.}


\colb{An EFST 
\beas
\clf\dmn \aea \z (\xi\dosw\n\m-\xi\dosw\m\n)
\eeas
derived from a conformal Killing vector $\xi$ also satisfies Maxwell's
equations; as before, $\z$ is a 
suitable dimensionful constant.  }

\colb{Noting that a conformal Killing vector
$\xi$ satisfies 
\beas
\xi\dosw\m\n + \xi\dosw\n\m \aea \dsf 2 d\lrr{\xi\upa{}\dn{;\a}} \gd\m\n
\eeas
where $d$ is the dimension of spacetime, 
the argument, presented earlier---for EFST derived from Killing vectors---can be easily adapted to show that $\clf\dmn$
satisfies the homogeneous Maxwell's equation
\beas
\clf\dosw {\a\b}\m+\clf\dosw {\m\a}\b +\clf\dosw{\b\m}\a \aea 0
\eeas
and also the  Maxwell's equation involving  source term
\bea 
\clf\umn\dso\n \aea \dsf{\clj\upm}{\ez c}, \qquad \clj\upm \deq  \ez c \z \lrs{
2 R\tu{\m\l}\xi\dnl + \dsf{2(d-1)}{d} (\xi\upb\dso\b)\tu{;\m}}
\leqn{ime}
\eea 
If $\xi$ is a Killing vector, then $\xi\upb\dso\b \deq 0$ and 
\ct{ime} reduces to \ct{maxe}.   It would be of interest to explore
the consequences of extending the KEF and KEC Postulates 
to 
conformal Killing vectors.}
\end{document}